\documentclass[aps,preprint,prb,floatfix,citeautoscript]{revtex4-1}
\usepackage{ulem}
\usepackage{dcolumn}
\usepackage{bm}
\usepackage{color}
\usepackage{amssymb}
\usepackage{amsmath}
\usepackage{graphicx}
\usepackage{amsfonts}
\usepackage{slashed}
\usepackage{pstricks}
\usepackage{float}
\usepackage{hyperref}
\usepackage{array}
\allowdisplaybreaks
\usepackage{dcolumn}
\usepackage{epsf}
\usepackage{soul}
\begin{document}
\title{Light-nuclei spectra from chiral dynamics}
\author{M.\ Piarulli$^{\rm a}$, A.\ Baroni$^{\rm b}$,  L.\ Girlanda$^{\rm c,d}$, A.\ Kievsky$^{\rm e}$, A.\ Lovato$^{\rm a,f}$,
Ewing Lusk$^{\rm g}$, L.E.\ Marcucci$^{\rm e,h}$, Steven C.\ Pieper$^{\rm a}$, R.\ Schiavilla$^{\rm b,i}$, M.\ Viviani$^{\rm e}$, and
R.B.\ Wiringa$^{\rm a}$}
\affiliation{
$^{\rm a}$\mbox{Physics Division, Argonne National Laboratory, Argonne, Illinois 60439, USA}\\
$^{\rm b}$\mbox{Department of Physics, Old Dominion University, Norfolk, Virginia 23529, USA}\\
$^{\rm c}$\mbox{Department of Mathematics and Physics, University of Salento, 73100 Lecce, Italy} \\
$^{\rm d}$\mbox{INFN-Lecce, 73100 Lecce, Italy} \\
$^{\rm e}$\mbox{INFN-Pisa, 56127 Pisa, Italy} \\
$^{\rm f}$\mbox{INFN-TIFPA Trento Institute of Fundamental Physics and Applications, 38123 Trento, Italy} \\
$^{\rm g}$\mbox{Mathematics and Computer Science Division, Argonne National Laboratory,} 
\mbox{Argonne, Illinois 60439, USA}\\
$^{\rm h}$\mbox{Department of Physics, University of Pisa, 56127 Pisa, Italy}\\
$^{\rm i}$\mbox{Theory Center, Jefferson Lab, Newport News, Virginia 23606, USA}
}
\date{\today}
%


\index{}\maketitle

{\bf 
A major goal of nuclear theory is to explain the spectra and stability of nuclei in terms
of effective many-body interactions amongst the nucleus' constituents---the nucleons, {\sl i.e.},
protons and neutrons.  Such an approach, referred to below
as the {\sl basic model} of nuclear theory, is formulated in terms of point-like nucleons,
which emerge as effective degrees of freedom, at sufficiently low energy, as a result of
a decimation process, starting from the fundamental quarks and gluons, described by
Quantum Chromodynamics (QCD).  A systematic way to account for the constraints imposed
by the symmetries of QCD, in particular chiral symmetry, is provided by chiral effective field
theory, in the framework of a low-energy expansion.  Here we show, in quantum Monte Carlo 
calculations accurate to $\leq\!2\%$ of the binding energy, that two- and three-body chiral 
interactions fitted {\sl only}
to bound- and scattering-state observables in, respectively, the two- and three-nucleon sectors,
lead to predictions for the energy levels and level ordering of nuclei in the mass range
$A\,$=$\,$4--12 in very satisfactory agreement with experimental data.  Our findings provide
strong support for the fundamental assumptions of the basic model, and pave the way to its
systematic application to the electroweak structure and response of these systems as well as
to more complex nuclei.
}

\vspace{1cm}
The nuclear Hamiltonian in the basic model is taken to consist of non-relativistic kinetic
energy, and two- and three-body interactions.  There are indications that four-body interactions
may contribute at the level of $\sim\,$100 keV in $^4$He, but current formulations of the basic
model do not typically include them (see, for example, Ref.~\cite{Carlson:2015}).  Two-body
interactions consist of a long-range component, for inter-nucleon separation $r \gtrsim 2$ fm,
due to one-pion exchange (OPE)~\cite{Yukawa:1935}, and intermediate- and short-range
components, for, respectively, $1 \,\,{\rm fm} \lesssim r \lesssim 2$ fm and $r \lesssim 1$ fm.
Up to the mid-1990's, such models were based almost exclusively on meson-exchange phenomenology.
The mid-1990's models~\cite{Stoks:1994,Wiringa:1995,Machleidt:2001} were constrained by fitting
nucleon-nucleon ($N\!N$) elastic scattering data up to lab energies of 350 MeV, with
$\chi^2$/datum $\simeq 1$ relative to the database available at the time~\cite{Stoks:1993}.
Two well-known and still widely used examples in this class are the Argonne $v_{18}$
(AV18)~\cite{Wiringa:1995} and CD-Bonn~\cite{Machleidt:2001}.
These so-called {\it realistic} interactions also contained
isospin-symmetry-breaking (ISB) terms.  At the level of accuracy required~\cite{Stoks:1993},
full electromagnetic interactions, along with strong interactions, had to be specified in order to
fit the data precisely, and the AV18 model included electromagnetic corrections up to order
$\alpha^2$ ($\alpha$ is the fine structure constant).

Already in the 1980's, accurate three-body calculations showed that contemporary $N\!N$
interactions did not provide enough binding for the three-body nuclei, $^3$H and $^3$He~\cite{Friar:1984}.
In the late 1990's and early 2000's this realization was extended to the spectra 
(ground and low-lying excited states) of light p-shell nuclei
in calculations based on quantum Monte Carlo (QMC) methods~\cite{Pudliner:1997}
and in no-core shell-model (NCSM) studies~\cite{Navratil:2000}.
Consequently, the basic model with $N\!N$ interactions fit to scattering data, without 
the inclusion of a three-nucleon ($3N$) interaction, is incomplete.

Because of the composite nature of the nucleon and, in particular,
the dominant role of the $\Delta$ resonance in pion-nucleon scattering, multi-nucleon
interactions arise quite naturally in meson-exchange phenomenology.  
The Illinois $3N$ interactions~\cite{Pieper:2001} contain a dominant two-pion exchange (TPE)---the 
venerable Fujita-Miyazawa interaction~\cite{Fujita:1957}---and smaller multi-pion exchange components
resulting from the excitation of intermediate $\Delta$'s.  The most recent version, Illinois-7 
(IL7)~\cite{Pieper:2008}, also contains phenomenological isospin-dependent central terms.  
The small number (four) of parameters that fully characterize it were determined, in conjunction 
with the AV18, by fitting 23 ground or low-lying nuclear states in the mass range $A$=3--10.  
The resulting AV18+IL7 Hamiltonian then led to predictions of about 100 ground- and excited-state 
energies up to $A$=12, including the $^{12}$C ground- and Hoyle-state energies, in good agreement
with the corresponding empirical values~\cite{Carlson:2015}.

A new phase in the evolution of the basic model, and renewed interest in its further development,
have been spurred by the emergence in the early 1990's of chiral effective field theory
($\chi$EFT)~\cite{Weinberg:1990,Weinberg:1991,Weinberg:1992}.  In $\chi$EFT the symmetries
of QCD, in particular its approximate chiral symmetry, are used to systematically
constrain classes of Lagrangians describing, at low energies, the interactions of baryons ($N$'s and
$\Delta$'s) with pions as well as the interactions of these hadrons with electroweak
fields~\cite{Park:1993,Park:1996,Park:2003}.  While the conventional meson-exchange
formulation of the basic model described earlier relied on an expansion in terms of exchanges
of heavier and heavier mesons (and hence shorter and shorter ranges of associated interactions),
the $\chi$EFT formulation has, by contrast, an expansion in powers of pion momenta as its
organizing principle, directly rooted in QCD.  From this perspective, it can be justifiably argued
to have put the basic model on a more fundamental basis, by providing a link between QCD
and its symmetries, and the strong and electroweak interactions in nuclei.

Within $\chi$EFT many studies have been carried out dealing with the construction of $NN$
and $3N$
interactions~\cite{Ordonez:1996,Epelbaum:1998,Entem:2003,Machleidt:2011,vanKolck:1994,Epelbaum:2002,Bernard:2011,Girlanda:2011,Krebs:2012} and accompanying ISB
corrections~\cite{Friar:1999,Friar:2004,Friar:2005}.
These interactions were typically formulated in momentum space, and included cutoff functions to
regularize their behavior at large momenta which, however, made them strongly non-local
when Fourier-transformed in configuration space, and therefore unsuitable for use with quantum
Monte Carlo methods~\cite{Carlson:2015}. Among these, in particular, Green's Function Monte Carlo (GFMC)
is the method of choice to provide reliable
solutions of the many-body Schr\"odinger equation---presently for up to $A\,$=12 nucleons---with
full account of the complexity of the many-body, spin- and isospin-dependent correlations induced by
nuclear interactions.

\section*{Nuclear Hamiltonian}
In order to overcome these difficulties, in recent years local, 
configuration-space chiral $NN$ interactions have been derived~\cite{Gezerlis:2013,Piarulli:2016}.
We will point out differences between these models below.   In the following, we focus on the
family of local interactions constructed by our group. 
They are written as the sum of an electromagnetic-interaction component, $v^{\rm EM}_{ij}$,
including first- and second-order Coulomb, Darwin-Foldy, vacuum polarization, and magnetic moment
terms (as in Ref.~\cite{Wiringa:1995}), and a strong-interaction component, $v_{ij}$, characterized
by long- and short-range parts~\cite{Piarulli:2016}.  The long-range part includes OPE and TPE terms up to
next-to-next-to-leading order (N2LO) in the chiral expansion~\cite{Piarulli:2015}, derived in the static limit from leading and
sub-leading $\pi N$ and $\pi N\Delta$ chiral Lagrangians.  Its strength is fully determined by the nucleon
and nucleon-to-$\Delta$ axial coupling constants $g_A$ and $h_A$, the pion decay amplitude
$f_\pi$, and the sub-leading low-energy constants (LECs, in standard notation) $c_1$, $c_2$, $c_3$,
$c_4$, and $b_3+b_8$ constrained by reproducing $\pi N$ scattering data.  In coordinate
space, this long-range part is represented by charge-independent central, spin, and
tensor components with and without isospin dependence ${\bm \tau}_i\cdot {\bm \tau}_j$
(the so-called $v_6$ operator structure),
and by charge-independence-breaking central and tensor components induced by OPE
and proportional to the isotensor operator
$T_{ij}\,$=$\,3\, \tau^z_i\,\tau^z_j -{\bm \tau}_i\cdot {\bm \tau}_j$.
The radial functions multiplying these operators are singular at the origin
(they behave as $1/r^n$ with $n$ taking on values up to $n\,$=$\, 6$), and
are regularized by a cutoff of the form~\cite{Piarulli:2015}
\begin{equation}
C_{R_{\rm L}}(r)= 1-\frac{1}{(r/R_{\rm L})^6 \, {\rm e}^{(r-R_{\rm L})/a_{\rm L}}+1} \ ,
\label{eq:cll}
\end{equation}
with $a_{\rm L}$ taken as $R_{\rm L}/2$, and the values for $R_{\rm L}$ considered here are given below.

The short-range part is described by charge-independent contact interactions,
specified by a total of 20 LECs---2 at LO, 7 at NLO, and 11 at
next-to-next-to-next-to-leading (N3LO)---and charge-dependent ones characterized
by 6 LECs---2 at LO, one each from charge-independence-breaking and
charge-symmetry-breaking (proportional, respectively, to $T_{ij}$ and
$\tau^z_i+\tau^z_j$), and 4 at NLO from charge-independence-breaking~\cite{Piarulli:2016}.
In the NLO and N3LO contact interactions, Fierz transformations have been utilized
to rearrange terms that in configuration space would otherwise lead to
powers of ${\bf p}$---the relative momentum operator---higher than two.
The resulting charge-independent interaction contains, in addition to the $v_6$ operator
structure, spin-orbit, ${\bf L}^2$ (${\bf L}$ is the relative orbital angular momentum),
and quadratic spin-orbit components, while the charge-dependent one retains central,
tensor, and spin-orbit components.  Both are regularized by multiplication
of a Gaussian cutoff,
\begin{eqnarray}
C_{R_{\rm S}}(r) &=& \frac{1}{\pi^{3/2}\, R_{\rm S}^3} {\rm e}^{-(r/R_{\rm S})^2} \ .
\label{eq:css}
\end{eqnarray}
These 26 LECs in the short-range part of $v_{ij}$ were constrained by a fit to the
$NN$ database, including the deuteron ground-state energy and two-neutron scattering
length, as assembled by the Granada group~\cite{Perez:2013}.   Two classes
of interactions were constructed, which only differ in the range of laboratory energy
over which the fits were carried out, either 0--125 MeV in class I or 0--200 MeV
in class II.  For each class, three different sets of cutoff radii $(R_{\rm S},R_{\rm L})$
were considered $(R_{\rm S},R_{\rm L})\,$=$\,(0.8,1.2)$ fm in set a, (0.7,1.0) fm in set b,
and (0.6,0.8) fm in set c.  The $\chi^2$/datum achieved by the fits in class I (II) was
$\lesssim 1.1(\lesssim1.4)$ for a total of about 2700 (3700) data points.  We will refer to these
high-quality $NN$ interactions generically as the Norfolk $v_{ij}$'s (NV2s), and designate those
in class I as NV2-Ia, NV2-Ib, and NV2-Ic, and those in class II as NV2-IIa, NV2-IIb, and NV2-IIc.

We observe that the models of Ref.~\cite{Gezerlis:2013} do not include $\Delta$ contributions
and only retain contact interactions up to NLO for a total of 9 LECs.  These LECs were fitted
to the phase-shifts of the Nijmegen partial-wave analysis~\cite{Stoks:1993}, rather than to
its database of $NN$ cross sections and polarization observables.  Some of the drawbacks
that this entails are discussed in Ref.~\cite{Piarulli:2015}.

\begin{center}
\begin{figure}[bth]
 \includegraphics[width=2.75in]{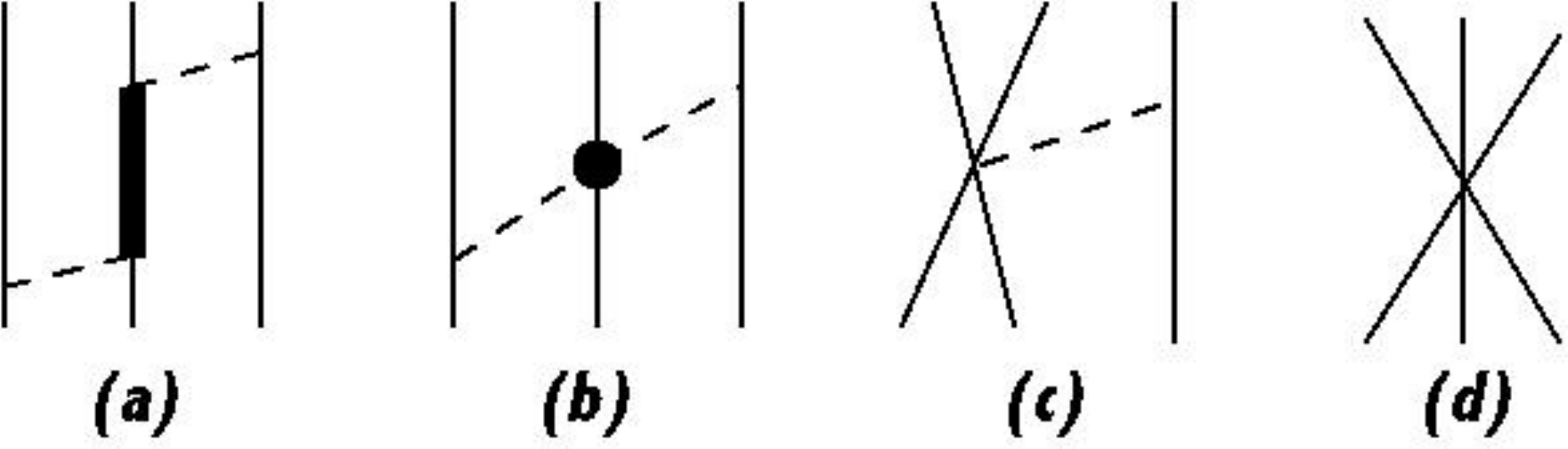}
 \caption{{\bf $3N$ chiral interaction.}
 Diagrams illustrating schematically the contributions to the $3N$ interaction.
 Nucleons, $\Delta$'s, and  pions are denoted by solid, thick-solid, and dashed
 lines, respectively.  The circle in panel (b) represents the vertex involving the LECs $c_1$, $c_3$,
 and $c_4$ in ${\cal L}^{(2)}_{\pi N}$.
}
\label{fig:f1}
\end{figure}
\end{center}
The NV2s were found to provide insufficient attraction, in GFMC calculations, 
for the ground-state energies of nuclei
with $A\,$=$\, 3$--6~\cite{Piarulli:2016,Lynn:2016,Lynn:2017}, thus corroborating the insight realized in the early
2000's within the older (and less fundamental) meson-exchange phenomenology.  To remedy
this shortcoming, we construct here the leading $3N$ interaction $V_{ijk}$ in $\chi$EFT,
including $\Delta$ intermediate states.  It is illustrated diagrammatically in Fig.~\ref{fig:f1}, and
consists~\cite{vanKolck:1994,Epelbaum:2002} of a long-range piece mediated by TPE and
denoted with the superscript $2\pi$, panels (a) and (b), and a short-range 
piece parametrized in terms of three contact interactions and denoted with the
superscript CT, panels (c) and (d),
\begin{equation}
V_{ijk}=\sum_{{\rm cyclic}\,\, ijk} \left(\,V^{2\pi}_{ijk}+V^{\rm CT}_{ijk}\,\right) \ .
\end{equation}
In configuration space, the TPE term from intermediate $\Delta$ states, panel (a) in Fig.~\ref{fig:f1},
and from interactions proportional to the LECs $c_1$, $c_3$, and $c_4$ in the sub-leading chiral Lagrangian
${\cal L}^{(2)}_{\pi N}$~\cite{Fettes:2000}, panel (b), reads
\begin{eqnarray}
\label{eq:e51}
V^{2\pi}_{ijk}&=&\frac{g_A^2}{256\, \pi^2} \frac{m_\pi^6}{f_\pi^4}\Big[
8\, c_1\, \Sigma_{ij}\, \Sigma_{kj} \, \mathcal{T}^{\,(+)}_{ijk}
+\,\frac{2}{9}\, \widetilde{c}_3 \, \Sigma^{(+)}_{ijk} \,\, \mathcal{T}^{\,(+)}_{ijk} \nonumber\\
&&-\frac{1}{9}\Big(\widetilde{c}_4+\frac{1}{4\, m}\Big) \Sigma^{(-)}_{ijk} \,\, \mathcal{T}^{\,(-)}_{ijk} \Big]\ , 
\end{eqnarray}
with spin and isospin operator structures defined, respectively, as
$\Sigma_{lm}\equiv \widetilde{Z}_\pi(r_{lm}) \, {\bm \sigma}_l\cdot\hat{\bf r}_{lm}$, where ${\bf r}_{lm}\equiv{\bf r}_l-{\bf r}_m$, and
\begin{eqnarray}
&&\Sigma^{(\mp)}_{ijk}\equiv \left[ \widetilde{X}_{ij} \, , \, \widetilde{X}_{jk} \right]_\mp ,
\,\,\,\, \mathcal{T}^{\,(\mp)}_{ijk} \equiv \left[{\bm \tau}_i \cdot {\bm \tau}_j \, , \, {\bm \tau}_j \cdot {\bm \tau}_k \right]_{\mp} ,\\
&&\qquad\widetilde{X}_{ij} \equiv \widetilde{T}_{\pi}(r_{ij})\, S_{ij} + \widetilde{Y}_{\pi}(r_{ij})\,
{\bm \sigma}_i \cdot {\bm \sigma}_j \ .
\end{eqnarray}
Here $\left [\, \dots\, ,\, \dots\right]_{\mp}$ denote commutators ($-$) or anti-commutators ($+$), $S_{ij}$ is the standard
tensor operator, ${\bm \sigma}_i$ and ${\bm \tau}_i$ are Pauli spin and isospin matrices relative to nucleon $i$, and
the regularized radial functions are defined as
\begin{eqnarray}
\widetilde{Y}_\pi(r)&=&\frac{e^{-{m_{\pi}}r}}{{m_{\pi}}r} \,C_{R_{\rm L}}(r) \ , \\
\widetilde{T}_{\pi}(r)&=&\left( 1 + \frac{3}{m_{\pi}\,r} + \frac{3}{m_{\pi}^2\, r^2} \right)  \widetilde{Y}_{\pi}(r) \ , \\
\widetilde{Z}_\pi(r)&=&-\left(1+\frac{1}{m_\pi\, r}\right)\widetilde{Y}_\pi(r) \ ,
\end{eqnarray}
where the cutoff $C_{R_{\rm L}}(r)$ is defined in Eq.~(\ref{eq:cll}).
Lastly, the LECs $\widetilde{c}_3$ and $\widetilde{c}_4$ are related to the corresponding
$c_3$ and $c_4$ in ${\cal L}^{(2)}_{\pi N}$ via
\begin{equation}
\widetilde{c}_3=c_3-\frac{h^2_A}{9\, m_{\Delta N}} \ ,\qquad \widetilde{c}_4=c_4+\frac{h^2_A}{18\, m_{\Delta N}} \ ,
\end{equation}
where $h_A$ and $m_{\Delta N}$ are, respectively, the $N$-to-$\Delta$ axial coupling constant and
$\Delta$-$N$ mass difference.  The values of these constants as well as the LECs $c_1$, $c_3$, and
$c_4$, the (average) pion mass $m_\pi$ and decay amplitude $f_\pi$, and (average) nucleon mass
$m$ and axial coupling constant $g_A$, are taken from Tables~I and~II of Ref.~\cite{Piarulli:2015}.

The CT term is parametrized as
\begin{eqnarray}
 V^{\rm CT}_{ijk}&=& \frac{g_A\, c_D}{96\, \pi} \, \frac{m_\pi^3}{\Lambda_\chi\, f_\pi^4}\,
 {\bm \tau}_i\cdot{\bm \tau}_k\, \widetilde{X}_{ik}\,  \left[\, C_{R_{\rm S}}(r_{ij}) +C_{R_{\rm S}}(r_{jk}) \,\right] \nonumber\\
&& +\frac{c_E}{\Lambda_\chi\, f_\pi^4}\, {\bm \tau}_i\cdot{\bm \tau}_k \, C_{R_{\rm S}}(r_{ij})\, C_{R_{\rm S}}(r_{jk})\ ,
\end{eqnarray}
where $C_{R_{\rm S}}(r)$ is the cutoff in Eq.~(\ref{eq:css}),
$\Lambda_\chi\,$ is the chiral-symmetry-breaking scale taken as $\Lambda_\chi\,$=$\,1$ GeV, and
the two (adimensional) LECs $c_D$ and $c_E$ are determined by simultaneously reproducing
the experimental $^3$H ground-state energy, $E_0(^3$H), and the neutron-deuteron ($nd$)
doublet scattering length, $^2a_{nd}$.  These observables are
calculated with hyperspherical-harmonics (HH) expansion methods.  By now, these variational methods
have achieved a high degree of sophistication (see below, and for a more extended recent review Ref.~\cite{Kievsky:2008}), permitting the accurate, virtually exact solution of the bound- and scattering-state
problem---the latter, both below and above two-body breakup thresholds---in the three- and four-nucleon systems.

The determination of $c_D$ and $c_E$ proceeds as follows.  For a range of
$c_D$ values we determine $c_E$ by reproducing either $E_0(^3$H) or
$^2a_{nd}$ (its central value).  The intercept of the resulting trajectories
in the $(c_D,c_E)$-plane provides the sought simultaneous solution.  This procedure is
repeated for each NV2-I(a-b) and NV2-II(a-b) with the cutoff radii $(R_{\rm S},R_{\rm L})$
in the Norfolk $3N$ interactions matching those of the corresponding
NV2s to make the NV2+3 models reported here; the $c_D,c_E$ values for each combination are listed in Table~\ref{tb:tb1}.
We observe that models NV2-Ic and NV2-IIc are not considered any further in the present work,
owing to the difficulty in the convergence of the HH expansion and the
severe fermion-sign problem in the GFMC imaginary-time propagation
with these interactions~\cite{Piarulli:2016}.
\begin{center}
\begin{table}[bth]
\caption{{\bf Fitted values of $c_D$ and $c_E$ and HH results for $A\,$=$\,$3--4 observables.}
The (adimensional) values of $c_D$ and $c_E$ obtained for the
different NV2+3 chiral interactions having
cutoff radii $(R_{\rm S},R_{\rm L})$ equal to (0.8,1.2) fm for models Ia and
IIa, and (0.7,1.0) fm for models Ib and IIb are shown along with the $^3$H, $^3$He, and $^4$He
ground-state energies (in MeV) and $nd$ doublet scattering length (in fm),
obtained in HH calculations without and with the inclusion of the three-body interactions;
the experimental values are  $E_0(^3$H)$\,$=$\,$--8.482 MeV, $E_0(^3$He)$\,$=$\,$--7.718 MeV, 
$E_0(^4$He)$\,$=$\,$--28.30 MeV, and $^2a_{nd}\,$=$\,(0.645\pm 0.010)$ fm.  The $E_0(^3$H)
and $^2a_{nd}$ observables are fitted when $3N$ interactions are included.}
 \begin{tabular}{c|rr|cccc|ccc}
        &   \multicolumn{2}{c}{} & \multicolumn{4}{c}{w/o $3N$} &
\multicolumn{2}{c}{with $3N$}\\
\hline
\hline
      Model & $c_D$ & $c_E$ &  $E_0(^3$H) & $E_0(^3$He) &  $E_0(^4$He)& $^2a_{nd}$&
       $E_0(^3$He) & $E_0(^4$He)\\
 \hline\hline
 Ia &   3.666 & --1.638 & --7.825 & --7.083 & --25.15 & 1.085 & --7.728 & --28.31  \\
 Ib & --2.061 & --0.982 & --7.606 & --6.878 & --23.99 & 1.284 & --7.730 & --28.31  \\
\hline                                                                                              
IIa &   1.278 & --1.029 & --7.956 & --7.206 & --25.80 & 0.993 & --7.723 & --28.17  \\
IIb & --4.480 & --0.412 & --7.874 & --7.126 & --25.31 & 1.073 &  --7.720 & --28.17  \\
\hline\hline
\end{tabular}
\label{tb:tb1}
\end{table}
\end{center}

\begin{center}
\begin{figure}[bth]
\includegraphics[width=17cm]{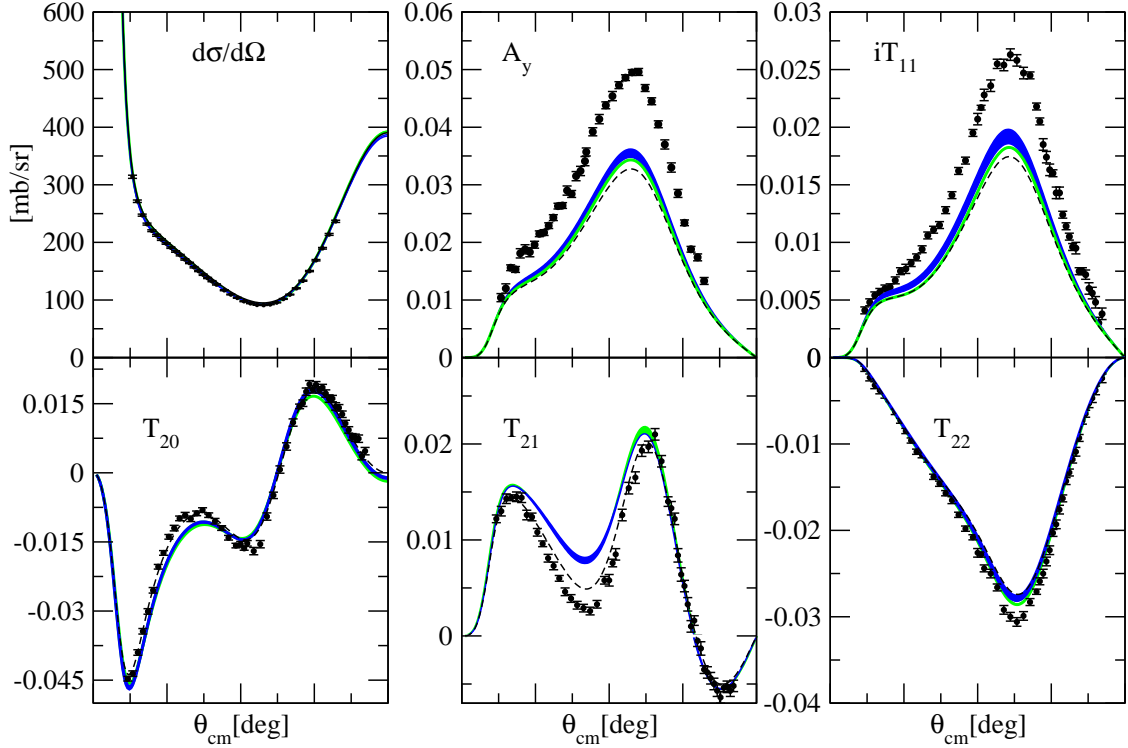}
\caption{{\bf Polarization observables in $pd$ elastic scattering at 3 MeV.}
Polarization observables obtained in HH calculations with the NV2+3 models Ia-Ib (IIa-IIb)
are shown by the green (blue) band. Also shown by the black dashed line are results obtained
with only the two-body interaction NV2-Ia.}
\label{fig:f2}
\end{figure}
\end{center}

In Table~\ref{tb:tb1} we also report the $nd$ scattering length and ground-state energies of
$^3$H, $^3$He, and $^4$He obtained without $3N$ interaction as well as those predicted for $^3$He
and $^4$He when this interaction is included (experimental values for the scattering length and
energies are taken, respectively, from Ref.~\cite{Schoen:2003} and Ref.~\cite{Audi:2003}).  Increasing
the laboratory-energy range over which the $NN$
interaction is fitted, from 0--125 MeV in class I to 0--200 MeV in class II, decreases the
$A\,$=$\,$3--4 ground-state energies calculated without the $3N$ interaction, by as much as 1.3 MeV
in $^4$He with model b.  However, when the $3N$ interaction is included, the effect is reversed
and much reduced, in $^4$He the increase amounts to 140 keV in going
from model Ib to IIb.  The dependence on the cutoff radii $(R_{\rm S},R_{\rm L})$, {\sl i.e.}, the
difference between the rows Ia-Ib and IIa-IIb, is significant without the $3N$ interaction, but
turns out to be negligible when it is retained, being in this case of the order of a few
keV and hence comparable to the numerical precision of the present HH methods.
This tradeoff is of course achieved through the large variation of the LECs $c_D$ and $c_E$,
which remain nevertheless of natural order in both models, a and b.

In Fig.~\ref{fig:f2} the differential cross section, and vector and tensor polarization
observables in proton-deuteron elastic scattering obtained with the present NV2+3 models
are compared to experimental data~\cite{Shimizu:1995}.  Theoretical
predictions remain essentially unchanged for Ia-Ib, but display a small variation
for IIa-IIb, as the cutoff radii $(R_{\rm S},R_{\rm L})$ are reduced from (0.8,1.2) fm in
models a  to (0.7,1.0) fm in models b.  The effect of the $3N$ interaction is small, marginally
improving (appreciably worsening) the agreement between theory and experiment for the
observables $A_y$, $i\, T_{11}$, and $T_{22}$ ($T_{20}$ and $T_{21}$).   In particular, the
well known discrepancy in the vector analyzing
power---the ``$A_y$ puzzle''~\cite{Gloeckle:1996}---persists.  It also appears to be unresolved when
higher-order chiral loops are accounted for in the long-range component of the $3N$ interaction~\cite{Golak:2014}.
Subleading contact terms in its short-range component, while having been formally derived~\cite{Girlanda:2011},
have yet to be implemented in calculations, since they depend on 10 unknown LECs.
Indeed, members of the present collaboration are currently involved in a fit of these LECs
to experimental data on $3N$ scattering observables.

\section*{Hyperspherical-harmonics expansion method}
The HH method uses hyperspherical-harmonics functions
as an expansion basis for the wave function of an $A$-body system~\cite{Kievsky:2008}.  In the
specific case of  $A\,$=$\,3$ and 4 nuclei, the bound-state wave function
 $\Psi_A$, having total angular momentum and parity quantum numbers $J^\pi$, is expanded as
 \begin{equation}
  \Psi_A=\sum_{[K_A]} u_{[K_A]}(\rho_A) {\cal B}_{[K_A]}(\Omega_A)\ ,
\label{eq1}
 \end{equation}
where ${\cal B}_{[K_A]}(\Omega_A)$ are fully
antisymmetrized HH-spin-isospin functions, which for three and four nucleons 
are characterized, respectively, by the set of quantum numbers $[K_3]\equiv [n_1,l_1,l_2,L,s,S,t,T]$ and
$[K_4]\equiv [n_1,n_2,l_1,l_2,l_3,l',L,s,s',S,t,t',T]$.
 The quantum numbers $n_i,l_i$ and $l'$ enter in the construction of the
HH vector and are such that the grand angular momenta are 
$K_3=2\,n_1+l_1+l_2$ and $K_4=2\,n_1+2\,n_2+l_1+l_2+l_3$.
The orbital angular momenta $l_i$ (and $l'$ for $A=4$) are
coupled to give the total orbital angular momentum $L$. The
total spin and isospin of the vector are indicated, respectively, by $S$ and $T$, 
and $s,s',t,t'$ denote intermediate couplings.

The hyperspherical coordinates $(\rho_A,\Omega_A)$ in Eq.~(\ref{eq1}) are given
by the hyperradius $\rho_A=(\sum_{i=1}^{A-1} {\bf x}_i^2)^{1/2}$ expressed in terms of
the $A$--$1$ Jacobi vectors ${\bf x}_i$ of the systems, and the hyperangles 
$\Omega_A=({\hat{\bf x}}_1 \ldots {\hat{\bf x}}_{A-1},\alpha_2\ldots\alpha_{A-1})$, with
${\hat{\bf{x}}}_i$ being the unit Jacobi vectors 
and $\alpha_i$ the hyperangular variables.
In the present application, the hyperradial functions 
are expanded in terms of generalized Laguerre polynomials multiplied by
an exponential function
\begin{equation}
u_\mu(\rho_A)= \sum_m C_{m,\mu}\;{\cal L}_m^{(3A-4)}(z) \; e^{-z/2}\ ,
\label{eq3}
\end{equation}
with $z=\beta\rho_A$, $\beta$ being a nonlinear parameter and 
$\mu\equiv[K_A]$.  After introducing the above expansion in Eq.~(\ref{eq1}), 
the wave function $\Psi_A$ is expressed compactly as
 \begin{eqnarray}
  \Psi_A&=&\sum_{m,\mu} C_{m,\mu}\; \Phi_{m,\mu}(\rho_A,\Omega_A) \ ,
\label{eqA} \\
\Phi_{m,\mu}(\rho_A,\Omega_A) 
&=& {\cal L}_m^{(3A-4)}(z) e^{-z/2} {\cal B}_{[K_A]}(\Omega_A)\ ,
 \end{eqnarray}
where the $\Phi_{m,\mu}(\rho_A,\Omega_A)$'s form a complete basis.

The ground-state energy $E_0$ follows from the Rayleigh-Ritz variational
principle.  This leads to a generalized eigenvalue problem, which is then
solved with standard numerical techniques~\cite{Kievsky:2008}.  The convergence
of the energy $E_0$ is studied in terms of the size of the basis.
For the three-nucleon system ($J^\pi\,$=$\,$1/2$^+$) all possible combinations of 
HH functions up to $l_1+l_2\,$=$\,$6 and isospin components $T\,$=$\,$1/2 and 3/2 have
been taken into account, thus attaining a level of accuracy of the order of a few
keV on the sought energy eigenvalue. For $A\,$=$\,4$ ($J^\pi\,$=$\,$0$^+$),
all possible combinations of HH functions up to $l_1+l_2+l_3\,$=$\,$6 ($l_1+l_2+l_3\,$=$\,$2)
having $T\,$=$\,$0 ($T\,$=$\,$1 and 2) have been considered, attaining in this case
a level of accuracy of about 20 keV for the $^4$He ground state energy~\cite{Viviani:2005}.
 
The $J\,$=$\,$1/2$^+$ $nd$ scattering wave function $\Psi_{nd}$, used to calculate
the doublet $nd$ scattering length, is expressed as $ \Psi_{nd}=\Psi_C+\Phi_{nd}$,
and $\Psi_C$ vanishes in the limit of large $nd$ separation.
It is expanded in terms of HH function and Laguerre polynomials as for the
bound state, using Eq.~(\ref{eqA}). The wave function
$\Phi_{nd}$ describes the system in the asymptotic region
\begin{eqnarray}
\Phi_{nd}&=&\sum_{{\rm cyclic}\,ijk}\sum_{L^\prime S^\prime}
[[s_i\otimes \phi_d(jk)]_{S^\prime}\otimes
Y_{L^\prime}({\hat{\bf{r}}_{nd}})]_{J J_z} \nonumber\\
&&\times [\delta_{L L^\prime}\delta_{S S^\prime} j_{L^\prime}(pr_{nd})
+ ^J\!\!R_{LS}^{L^\prime S^\prime}(p) n_{L^\prime}(pr_{nd}) g(pr_{nd})]
\label{eq:phi} \ , 
\end{eqnarray}
where $\phi_d$ is the deuteron wave function, $p$ ($r_{nd}$) is the $nd$ relative momentum (distance),
and $j_L$ ($n_L$) are the regular (irregular) Bessel functions.  The function $g(pr_{nd})$ modifies $n_L$ at small
$r_{nd}$ by regularizing it at the origin, and $g(pr_{nd})\rightarrow 1$
for $r_{nd}\ge 10$ fm. Finally, the real parameters $^JR_{LS}^{L^\prime S^\prime}(p)$ 
are the $R$-matrix elements which determine phase shifts and, for coupled channels, mixing angles.

The unknown quantities in $\Psi_{nd}$, {\sl i.e.}, the coefficients $C_{m,\mu}$ in the expansion
of $\Psi_C$ and $R$-matrix elements $^JR_{LS}^{L^\prime S^\prime}(p)$ in $\Phi_{nd}$, are obtained by utilizing
the Kohn variational principle, which leads to a set of inhomogeneous coupled equations for $C_{m,\mu}$ and
a set of algebraic equations for $^JR_{LS}^{L^\prime S^\prime}(p)$, solved by standard
techniques~\cite{Kievsky:2008}.  In particular, the $nd$ doublet scattering length simply follows from
$^2a_{nd}=-\lim_{p\rightarrow 0} {^{1/2}R_{0\, 1/2}^{0\, 1/2}(p)}$, and
the convergence of the HH expansion for $^2a_{nd}$ is established with a procedure
similar to that outlined above for the bound state, ultimately achieving an accuracy of the order of 0.001 fm.
The extension of the method to describe proton-deuteron scattering, specifically the
inclusion, in the asymptotic channels, of the Coulomb interaction (and higher-order electromagnetic interactions
as retained in the NV2 models), is discussed in Ref.~\cite{Kievsky:2001}.
  
\section*{Quantum Monte Carlo methods}
For the NV2+3-Ia model, we have calculated the energies for $\sim\!100$ 
nuclear states in $A$=6--12 nuclei using quantum Monte Carlo methods.  
A subset of these spectra calculations is shown in Fig.~\ref{fig:f3} and
compared to QMC results for the phenomenological AV18+IL7 Hamiltonian 
and to experiment.
The QMC method is briefly described below; a more complete description is 
given in Refs.~\cite{Piarulli:2016,Carlson:2015}. 

The QMC calculation for a given nuclear state is made in two steps:
(i) a variational Monte Carlo (VMC) calculation, in which a trial wave 
function is optimized by minimizing its energy expectation value, and 
(ii) a GFMC calculation, which filters out 
excited state contamination in the trial wave function by a propagation in 
imaginary time, to project out the lowest-energy wave function of given 
quantum numbers.
Energy calculations have a statistical error and some well-controlled 
systematic errors $\sim\,$(1--2)\% of the binding energy.

The VMC trial wave function $\Psi_T$ is constructed to be explicitly antisymmetric
and translationally invariant, with quantum numbers ($J^\pi;T$) of the state
of interest, where $T$ is the total isospin.
It is built up from a product of one-, two-, and three-body correlations
that have space, spin and isospin dependence induced by the Hamiltonian.
The $\Psi_T(J^\pi;T)$ is represented as a vector in spin-isospin space with 
order $2^A \left( ^A_Z \right)$ components, each of which is a function in
$3A$-dimensional configuration space.
It has a total of 50--100 variational parameters which are optimized to
give the lowest upper bound to the many-body energy expectation value, 
\begin{equation}
\label{eq:energy_vmc}
 E_T=\frac{\langle \Psi_T|H|\Psi_T\rangle}{\langle \Psi_T|\Psi_T\rangle}\geq E_0 \ ,
\end{equation}
where the quadrature is evaluated by a Metropolis Monte Carlo algorithm.
The search for optimal parameters (many of which do not vary greatly from
nucleus to nucleus) is made with the aid of automated search routines.
For $A$=6--12 nuclei, there can be multiple states with the same $(J^\pi;T)$
quantum numbers, {\sl e.g.}, three $1^+$ p-shell states in $^6$Li; we build
complete sets of orthogonal $\Psi_T$ for these nuclei.

The $\Psi_T$ serves as the starting point of a GFMC calculation, which projects
out the lowest energy state $\Psi_0$ with the same quantum numbers
by the evolution in imaginary time $\tau=-i\, t$:
\begin{equation}
|\Psi_0\rangle \propto \lim_{\tau\to\infty}|\Psi(\tau)\rangle=\lim_{\tau\to\infty}e^{-(H-E_0)\,\tau}\,|\Psi_T\rangle\ .
\end{equation}
The GFMC propagator ${\rm exp}[-(H-E_0)\,\tau]$ is evaluated stochastically in
small time steps $\Delta\tau$ with $\tau\,$=$\,n\,\Delta \tau$, and in 
practice is made with a simplified version $H^\prime$ of the Hamiltonian,
the small difference $\langle H-H^\prime \rangle$ being evaluated
perturbatively.  In calculations that are performed with a three-nucleon
potential, the $H^\prime$ is modified to make $\langle H-H^\prime \rangle \sim 0$;
however, such capability does not exist for calculations with only two-nucleon interactions.

The desired expectation values of ground-state and low-lying excited state
observables are then computed approximately by 
\begin{equation}
\label{eq:extrap}
 \langle{\mathcal O}(\tau)\rangle\equiv\frac{\langle\Psi(\tau)|{\mathcal O}|\Psi(\tau)\rangle}{\langle\Psi(\tau)|\Psi(\tau)\rangle}\approx \langle{\mathcal O}(\tau)\rangle_{\rm M} +
[\langle{\mathcal O}(\tau)\rangle_{\rm M}-\langle{\mathcal O}\rangle_{\rm V}]\ ,
\end{equation}
where $\langle{\mathcal O}\rangle_{\rm V}$ is the variational expectation 
value and $\langle{\mathcal O}\rangle_{\rm M}$ is the ``mixed'' estimate
\begin{equation}
\label{eq:mme}
\langle{\mathcal O}(\tau)\rangle_{\rm M}= \frac{\langle\Psi(\tau)|{\mathcal O}|\Psi_T\rangle}{\langle\Psi(\tau)|\Psi_T\rangle}\, .
\end{equation}
For the specific case ${\mathcal O} = H^\prime$ the mixed estimate is exactly
equivalent to $\langle{\mathcal O}(\tau/2)\rangle$ and the GFMC propagation
provides a convergent upper bound.
Energies ordinarily converge very rapidly in $\tau$ and the final answer 
with its statistical error is taken as the average over the $\tau \geq 0.1$ 
MeV$^{-1}$ points, typically up to
a maximum $\tau \sim 0.3$ or 0.4 MeV$^{-1}$.

As in QMC applications for other systems, such as for those in condensed matter, there is a 
well-known fermion sign problem due to the accumulation of bosonic noise
during the GFMC propagation, which gets worse with increasing system size.
The desired fermionic component is projected out by the antisymmetric $\Psi_T$
in the mixed estimate, but a constraint must be placed on the propagation
to keep statistical noise from overwhelming the fermion signal.
The constraint can be relaxed for the last 10--40 propagation time steps to
reduce a possible systematic error before the statistical error grows too much.
The fermion sign problem is much worse for the NV2+3 interactions than
for the older AV18+IL7 Hamiltonian; 
constrained path propagation is needed even for $A$=4 as opposed to 
$A$=7 with AV18+IL7.  To reduce the sign problem, we have used a
propagation time step $\Delta\tau = 0.00025$ MeV$^{-1}$, with
expectation values being evaluated after every 80 propagation steps
as opposed to the $\Delta\tau = 0.0005$ MeV$^{-1}$ that is used
for AV18+IL7 Hamiltonian.

For higher excited states of the same $(J^\pi;T)$, the GFMC might not be
expected to avoid mixing in some of the lowest energy state and thus
obtaining excitation energies that are too low.
However, with orthogonal starting $\Psi_T$, the GFMC propagation tends to
preserve orthogonality very well, and explicit corrections can be made so
that the overlap between different wave functions vanishes within statistical 
errors.  Many tests of the correctness of GFMC results have been made;
the extracted eigenenergies are reliable to better than 2\%~\cite{Carlson:2015}.

The computational requirements for this QMC method grow exponentially with
the number $A$ of nucleons, so while a four-body calculation is suitable
for a desktop machine, the final $^{12}$C ground state calculation reported here required
650,000 cpu-hours on the massively parallel Theta supercomputer (3,624 Intel 
Knight's Landing nodes with 64 cpus/node) of the Argonne Leadership Computing 
Facility.
The QMC codes are written in \textsc{fortran} and use \textsc{mpi} and 
\textsc{openmp} for parallelization.
While Monte Carlo calculations are often thought of as ``embarassingly
parallel", the GFMC propagation involves killing and replication of 
configurations which could lead to significant inefficiencies in a parallel
environment.
Also, for the largest nuclei, the calculation of a single Monte Carlo sample
must be spread over many nodes.
For these reasons the Asynchronous Dynamic Load Balancing 
(\textsc{adlb}) library and the Distributed MEMory (\textsc{dmem}) library,
which operate under \textsc{mpi}, were developed for our calculations~\cite{Lusk}.

\section*{Nuclear spectra: Theory confronts experiment}

Before presenting the GFMC predictions for the spectra of larger nuclei, it is
worthwhile comparing the HH and GFMC results for the three- and four-nucleon
bound states.  The GFMC-calculated ground-state energies with model NV2+3-Ia
are  $E_0$($^3$H)\,=\,--8.463(9), $E_0$($^3$He)\,=\,--7.705(9), and $E_0$($^4$He)\,=\,--28.24(3),
where the Monte Carlo statistical errors are given in parentheses.  The small differences
($\lesssim 0.5\%$) between the HH results listed in Table~\ref{tb:tb1} and the GFMC
ones are due in part to intrinsic numerical inaccuracies of these methods, and in part
to the fact that the HH wave functions include small admixtures with total isospin
$T\,$=$\,$3/2 for $A\,$=3 nuclei, and $T\,$=$\,$1 and 2 for $A\,$=4, beyond their
corresponding dominant isospin components with $T\,$=$\,$1/2 and $T\,$=$\,$0.
These admixtures are induced by ISB terms present in the NV2 interaction models,
which are neglected in the GFMC calculations.  The associated systematic
error, however, is quite small; for example, an HH calculation ignoring
these admixtures in $^4$He finds a reduction in binding of about 19 keV,
hence within the numerical noise of the HH method itself.

\begin{center}
\begin{figure}[bth]
\includegraphics[width=18cm]{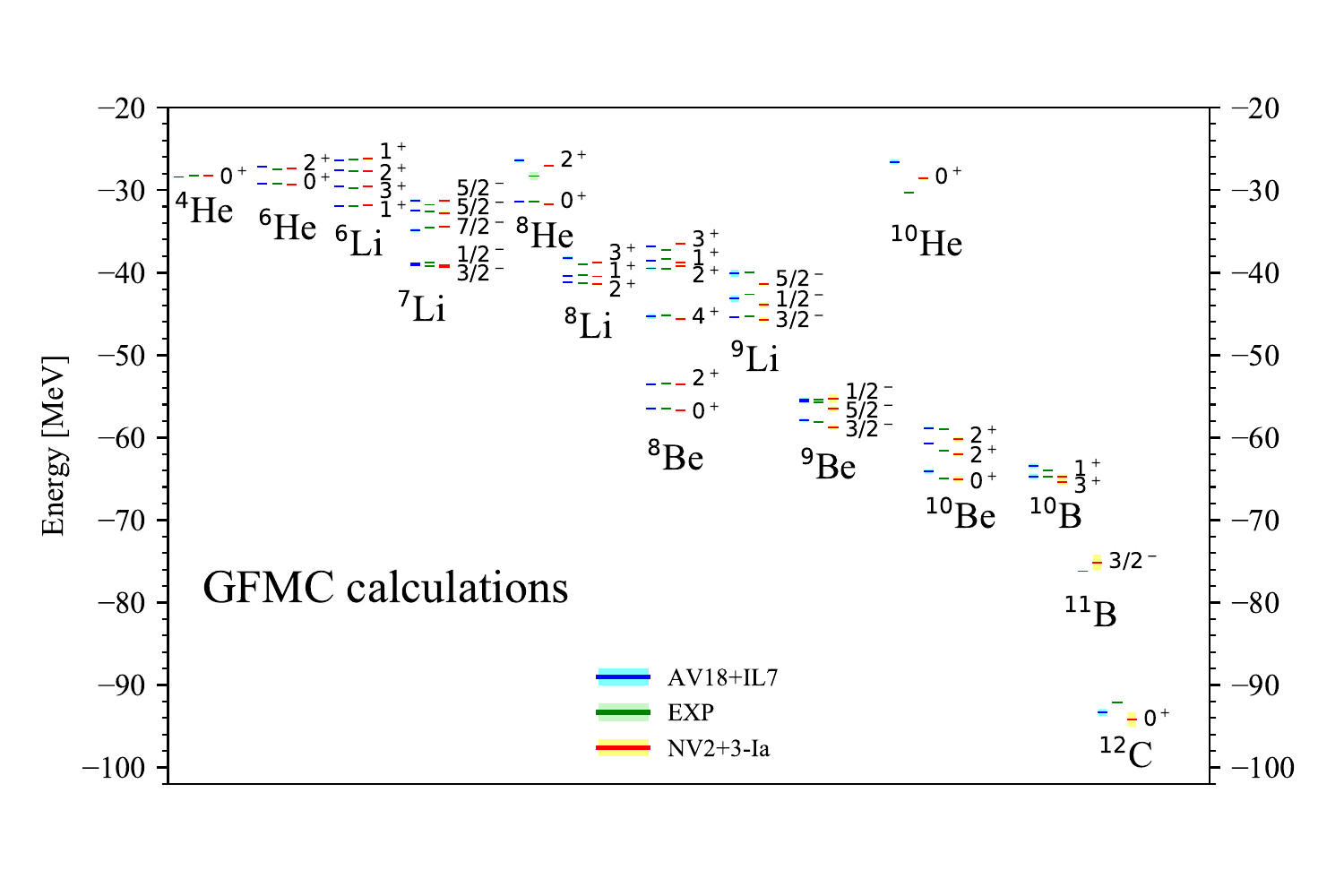}
\caption{{\bf Spectra of $A\,$=$\,$4--12 nuclei.}  The energy spectra
obtained with the NV2+3-Ia chiral interactions are compared to experimental data.
Also shown are results obtained with the phenomenological AV18+IL7 interactions.}
\label{fig:f3}
\end{figure}
\end{center}

The GFMC energy results calculated with the NV2+3-Ia model are shown in
Fig.~\ref{fig:f3} for 37 different nuclear states in $A$=4--12 nuclei.
They are compared to results from the older AV18+IL7 model and experiment.
The agreement with experiment is impressive for both Hamiltonians, with
absolute binding energies very close to experiment, and excited states
reproducing the observed ordering and spacing, indicating reasonable
one-body spin-orbit splittings.
The rms energy deviation from experiment for these states is 0.72 MeV
for NV2+3-Ia compared to 0.80 MeV for AV18+IL7 (note that $^{11}$B has not been
computed with AV18+IL7). The signed average deviations, +0.15 and --\,0.23 MeV
respectively, are much smaller; indicating no systematic over-
or under-binding of the Hamiltonians.
For both Hamiltonians, the inclusion of the  $3N$ interactions is 
in many cases necessary to get ground states that are correctly bound against
breakup.  For example, $^6$He is not bound with just the $N\!N$ interaction~\cite{Piarulli:2016},
but is in the current work.
The lowest $3^+$ and $1^+$ states of $^{10}$B are of particular interest.
For both AV18 and NV2-Ia without $3N$ interactions, the $1^+$ state is
incorrectly predicted as the ground state (for NV2-Ia by 1.9 MeV) but
including the $3N$ interactions gives the correct $3^+$ ground state.
However, it is important to emphasize that in the AV18+IL7 model 
the four parameters in the $3N$ interaction are fitted to the energies
of many nuclear levels up to $A=10$.

Twelve of the states shown are stable ground states, while another six
are particle-stable low-lying excitations, {\sl i.e.}, they decay only by
electroweak processes.
The remaining states are particle-unstable, {\sl i.e.}, they can decay by nucleon
or cluster emission, which is much more rapid than electroweak decay, but about 
half of these have narrow decay widths $\leq 100$ keV.
Because GFMC does not involve any expansion in basis functions, it
correctly includes effects of the continuum.  This means that if
the propagation is continued to large enough imaginary time, the
wave function will evolve to separated clusters and the energy
to the sum of the energies of those clusters.
For the physically narrow states, however, the GFMC propagation starting
from a confined variational trial function reaches
a stable energy without any noticeable decay over the finite $\tau$ used
in the present calculations, and this is the energy we quote.
For physically very wide states ($> 1$ MeV) this decay is observed in
the calculations, {\sl e.g.}, in the first $2^+$ and $4^+$ states in $^8$Be, as a 
smooth energy decline beyond $\tau \sim 0.1$ MeV$^{-1}$~\cite{Pastore:2014}.
In such cases, the rms radius also shows a smooth growth, indicative that
the propagation is disassembling the system into its component parts.
In these few cases the energy of the state is estimated from the value at
the beginning of the smooth energy decline.
Additional particle-stable isobaric analog states, {\sl e.g.}, in $^8$B and 
$^{9,10}$C, have been calculated in GFMC, but are not shown.

A VMC survey of more than 60 additional states has also been made, including
higher excited states, more isobaric analog states, {\sl e.g.}, in $^7$Be, and
various particle-unstable nuclei like $^7$He, $^8$C, and $^9$B.
For the nuclei shown in Fig.~\ref{fig:f3}, the VMC trial functions underbind
$^4$He by 1 MeV, and miss 1--1.5 MeV/nucleon binding in the $A \ge 6$ nuclei.
However, they get the same ordering of excitations as the final GFMC 
calculations, with very similar energy splittings.
The VMC survey of additional states indicates a continued good agreement
with known states.
While the most important test of a Hamiltonian is the ability to reproduce
known states, it is also important to {\it not} predict states in places where
they are not observed, {\sl e.g.}, predicting a particle-stable $^{10}$He
ground state would be a failure of the model.
The VMC survey has found no such problems for either the NV2+3-Ia 
or AV18+IL7 models.

\section*{Conclusions and future work}

The very satisfactory agreement between the predicted and
observed spectra validates the present formulation of the basic
model in terms of $NN$ and $3N$ chiral interactions, constrained
by data in the two- and three-nucleon systems only.  Key to this
significant advance is our group's ability to reliably solve the nuclear many-body
problem for bound states of up to $A\,$=$\,$12 nuclei with QMC
methods, and for the three- and four-nucleon bound and scattering
states with HH methods.  This capability, particularly for QMC, is
driven by ever expanding computational resources and by
continuing improvements in algorithms.  In a broader context,
the basic model developed here justifies the
program of nuclear theory aimed at understanding the
structure and reactions of nuclei solely on the basis 
of two- and three-nucleon forces.

In future, we plan to calculate the nuclear spectra for other
models---indeed, calculations with NV2+3-IIb have already begun---and
to refine the $3N$ chiral interaction by retaining subleading contact terms.
We will also be studying other nuclear properties, such as magnetic moments
and electroweak transitions, for these Hamiltonians.
An initial VMC survey of numerous $M1$ and $E2$ electromagnetic
and Gamow-Teller (GT) weak transitions finds very similar matrix elements
for NV2+3-Ia and AV18+IL7 when there is a large spatial overlap between
the initial and final states; this means reasonable agreement with
experiment since the AV18+IL7 model gives fairly good results in
such cases~\cite{Pastore:2014,Pastore:2013}.
However, when the transitions proceed from what is a large spatial symmetry
component in the initial state to a small component in the final state, the
NV2+3-Ia model often produces significantly larger matrix elements, which may
potentially lead to better agreement with experiment.
An essential aspect of this future work is the development of 
two-body electroweak currents consistent with the new models---in particular,
the two-body currents have been shown to make major contributions to 
magnetic moments and $M1$ transitions with the AV18+IL7 Hamiltonian~\cite{Pastore:2014,Pastore:2013}.

\section*{Acknowledgments}
The work of~M.P.,~A.L.,~E.L.,~S.C.P., and~R.B.W has been supported by 
the NUclear Computational Low-Energy Initiative (NUCLEI) SciDAC
project. This research is further supported by the U.S.~Department of
Energy, Office of Science,  Office of Nuclear Physics, under contracts
DE-AC02-06CH11357 (M.P.,~A.L.,~S.C.P., and~R.B.W.) and DE-AC05-06OR23177
(R.S.).  It used computational resources provided by 
Argonne's Laboratory Computing Resource Center, 
by the Argonne Leadership Computing Facility, which is a DOE Office of Science User Facility 
supported under Contract DE-AC02-06CH11357 (via a Theta Early Science grant),
and by the National Energy Research Scientific Computing Center (NERSC).
\end{document}